\documentclass[prl,twocolumn,aps]{revtex4-2}
\usepackage{amsmath}
\usepackage{amssymb}
\usepackage{amsthm}
\usepackage{graphicx}
\usepackage{bm}
\usepackage{physics}
\usepackage{mathtools}
\usepackage{xcolor}
\usepackage[caption=false]{subfig}

\usepackage{afterpage}

\newtheorem*{theorem*}{Theorem}

\usepackage{hyperref}
\definecolor{myblue}{RGB}{0,0,255}
\hypersetup{
    colorlinks,
    citecolor=myblue,
    linkcolor=myblue,
    urlcolor=myblue
}

\usepackage{subfig}
\usepackage{stmaryrd}
\usepackage{multirow}

\usepackage{booktabs}
\usepackage{float}

\begin{document}

\title{Energy Inference of Black-Box Quantum Computers Using Quantum Speed Limit}
\author{Nobumasa Ishida}
\email{ishida@biom.t.u-tokyo.ac.jp}
\affiliation{Department of Information and Communication Engineering, Graduate School of Information Science and Technology, The University of Tokyo, Tokyo 113-8656, Japan}
\author{Yoshihiko Hasegawa}
\email{hasegawa@biom.t.u-tokyo.ac.jp}
\affiliation{Department of Information and Communication Engineering, Graduate School of Information Science and Technology, The University of Tokyo, Tokyo 113-8656, Japan}

\begin{abstract}
Cloud-based quantum computers do not provide users with access to hardware-level information such as the underlying Hamiltonians, which obstructs the characterization of their physical properties. We propose a method to infer the energy scales of gate Hamiltonians in such black-box quantum processors using only user-accessible data, by exploiting quantum speed limits. Specifically, we reinterpret the Margolus--Levitin and Mandelstam--Tamm bounds as estimators of the energy expectation value and variance, respectively, and relate them to the shortest time for the processor to orthogonalize a quantum state. This shortest gate time, expected to lie on the nanosecond scale, is inferred from job execution times measured in seconds by employing gate-time amplification. We apply the method to IBM's superconducting quantum processor and estimate the energy scales associated with single-, two-, and three-qubit gates. The order of estimated energy is consistent with typical drive energies in superconducting qubit systems, suggesting that current gate operations approach the quantum speed limit. Our results demonstrate that fundamental energetic properties of black-box quantum computers can be quantitatively accessed through operational time measurements, reflecting the conjugate relationship between time and energy imposed by the uncertainty principle.
\end{abstract}

\maketitle

\textit{Introduction---}Uncertainty principle reveals the fundamental connection between conjugate physical quantities in terms of the possibility of simultaneous measurements. Among the most significant, the energy-time uncertainty is manifested as the quantum speed limit (QSL) \cite{Mandelstam1945-mv,Margolus1998-pq,Levitin2009-fz,Deffner2013-mv,Taddei2013-fg,Caneva2009-zf, Deffner2017-hz,Campaioli2017-rd,Mohan2022-do}. It dictates the minimum time for a quantum state to evolve into an orthogonal state in terms of the system energy, serving as a design principle in quantum technologies, including quantum computation \cite{Deffner2020-vt}, communication \cite{Yung2006-mx,Murphy2010-ne}, and sensing \cite{Herb2024-ez}. A common premise in the literature, however, is that both conjugate observables are accessible to experimenters. In practice, there exist situations in which access is one-sided, namely one of the two observables is not accessible at all.

\begin{figure}[t]
    \centering
    \includegraphics[width=0.8\linewidth]{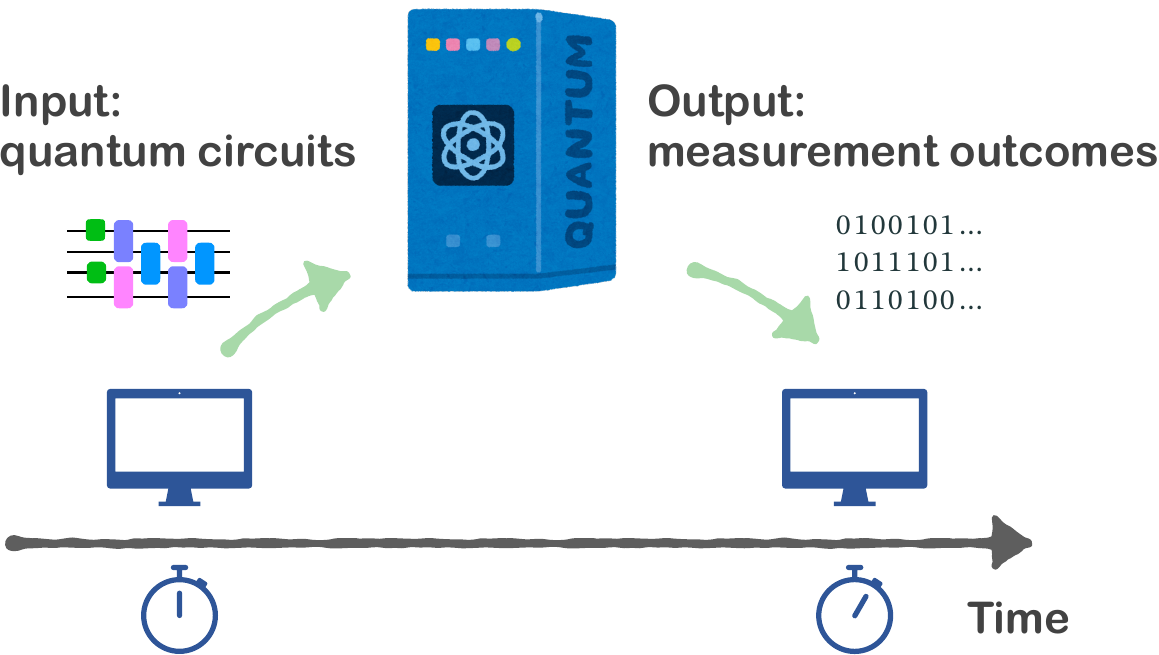}
    \caption{Schematic of the black-box quantum computer. The user can submit quantum circuits and receive measurement outcomes along with the execution time of the whole job. The internal structure of the quantum computer, such as the Hamiltonian generating the dynamics, is hidden from the user.}
    \label{fig:settings}
\end{figure}

Cloud-based quantum computers are an extreme example (Fig.~\ref{fig:settings}). A user can submit a quantum circuit and obtain measurement outcomes from many shots and the execution time of the entire job, but cannot access hardware information, such as qubit modality, pulse shapes, energy gaps, and drive strengths \cite{Nguyen2024-zu}. Nevertheless, these parameters characterizing their Hamiltonians are indispensable from the viewpoint of device physics, including scientific reproducibility, cooling loads \cite{Krinner2019-ke}, and understanding error mechanisms. What is the operational meaning of uncertainty relations in such a one-sided setting?

We provide a general method for inferring the internal energy scales of the black-box quantum computers using the QSL. The key idea is that knowing time is knowing energy, as the QSL filters a possible energy given time. Specifically, we measure the fastest time $\tau$ that quantum computers can orthogonalize a state and estimate the expectation value and variance of the qubit driving Hamiltonian using the inverse of the Margolus--Levitin (ML) \cite{Margolus1998-pq} and Mandelstam--Tamm (MT)  \cite{Mandelstam1945-mv} bounds. We show that $\tau$ (on the order of nanoseconds) can be extracted from the job execution time of a quantum computer (on the order of seconds) by the method called gate-time amplification. Therefore, our approach requires only the user-accessible time information and is universally applicable: it is platform-independent, valid for time-dependent Hamiltonians, and applicable to any number of energy levels.

We apply the method to IBM's superconducting quantum computer \texttt{ibm\_torino} \cite{IBMUnknown-to} and estimate the energy expectation values and variances of the Hamiltonians for single-, two-, and three-qubit gates. We obtain lower bounds of order $10^{-27}$ J for single- and two-qubit gates. It is consistent with the order of the typical driving energies in superconducting-qubit experiments. This result not only highlights the validity of the method but also suggests that the actual gates in the quantum processor operate near the QSL \cite{Negirneac2021-ua,Hegde2022-mv,Howard2023-hb}. Moreover, for three-qubit gates, we obtain a lower bound of order $10^{-28}$ J as the time-averaged energy of the effective three-qubit dynamics. This study provides a viewpoint to infer otherwise inaccessible Hamiltonian information through the uncertainty principle and opens a route to evaluate physical properties of black-box quantum devices.

\textit{Settings---}We consider a gate-based quantum computer that behaves as a black box for the user. We assume that the user can perform only the following operations.
(i) Submit a job specifying quantum circuits and the number of shots $N_{\rm shots}$.
(ii) After job execution, the measurement outcomes of each shot and the total job execution time $T_{\rm exec}$ are obtained.
The user cannot access any information about the internal hardware, including the physical implementation of qubits (superconducting, trapped ions, etc.), native gate sets, detailed pulse shapes, and the qubit energy gaps and the strength of the driving Hamiltonian. We assume, however, that the quantum computer performs accurate qubit operations as designed by the input circuits. We note that the small imperfections in physical implementations are not relevant in the following.

Inside the quantum computer, time evolution on the $n$-qubit computational basis $\{\ket{0},\ket{1}\}^{\otimes n}$ is realized by a (generally time-dependent) Hamiltonian $H_n(t)$. We focus on the state change on the basis, and thus $H_n(t)$ is in a rotating frame eliminating irrelevant global phases and rapidly oscillating components. We do not assume the specific form of $H_n(t)$, such as whether it is time-independent or time-dependent. In addition, the dynamics is not necessarily restricted to the computational subspace during gate operations; higher levels (e.g., $\ket{2}$) may participate, as in the undesired leakage  \cite{Hyyppa2024-au} or the CZ gate via the $|02\rangle$ state \cite{Rol2019-oo} in transmons. While we focus on the case where the computational subspace has two levels, the method is applicable to qudit systems \cite{Cho2024-qk} as well.

To quantify the energy scale of the evolution generated by $H_n(t)$, we define the time-averaged energy $E_n$ and energy variance $\Delta E_n$ for an $n$-qubit state $\ket{\psi(t)}$ as
\begin{align}
    E_n = \frac{1}{T}\int_0^T E_n(t)dt,\quad \Delta E_n = \frac{1}{T}\int_0^T \Delta E_n(t)dt,
\end{align}
where $T$ is the total time for the operation, $E_n(t) = \bra{\psi(t)} H_n(t) \ket{\psi(t)}$ is the instantaneous energy expectation, and $\Delta E_n(t) = \sqrt{\bra{\psi(t)} H_n^2(t) \ket{\psi(t)} - E_n^2(t)}$ is the instantaneous energy variation.

Such driving energies are relevant for understanding the performance and thermal properties of quantum computers. For example, drive energy characterizes the heat loads on refrigeration systems in superconducting qubits: to supply sufficient drive power to qubits, the total power of control signals must be large enough to compensate for about 60 dB of attenuation from room temperature to the millikelvin stage \cite{Krinner2019-ke}. The associated heat dissipation is of interest in quantum thermodynamics and assessing energy advantages in quantum computing \cite{Blok2025-fh,Auffeves2022-nh,Meier2025-kz}.

\textit{Methods---}QSLs were originally introduced as relations that constrain time by energy. The MT bound provides the minimum time $T_{\rm QSL}^{\rm MT}$ required for a pure state $\ket{\psi(t)}$ to reach a state orthogonal to the initial state: $T_{\rm QSL}^{\rm MT} = \pi\hbar/(2 \Delta E)$, where $\Delta E$ is the energy variance. On the other hand, when the Hamiltonian is time-independent and the ground-state energy is defined as zero, the ML bound provides another limit in terms of the mean energy $E$: $T_{\rm QSL}^{\rm ML} = \pi\hbar/(2 E)$. In general, the effective QSL is given by the greater of these two bounds \cite{Levitin2009-fz,Ness2021-ks}.

In this work, we reinterpret these relations as a means to estimate the energy scale of the Hamiltonian from time. We highlight the fact that if the state evolves to an orthogonal state within time $T$, then the evolution must satisfy $T \ge T_{\rm QSL}$, and thus
\begin{align}
    E \geq \frac{\pi\hbar}{2 T},\quad \Delta E \geq \frac{\pi\hbar}{2 T}
    \label{eq:inverse_QSL}
\end{align}
holds. Therefore, by investigating the fastest operation that orthogonalizes a state in a quantum computer, one can infer lower bounds on the mean energy and variance of the corresponding Hamiltonian.

From this viewpoint, we define the following energy estimators for $n$-qubit gates:
\begin{align}
    E_{n}^{\rm est} = \frac{\pi\hbar}{2 \tau_n},\quad \Delta E_{n}^{\rm est} = \frac{\pi\hbar}{2 \tau_n}, \label{eq:estimator_multi}
\end{align}
where 
\begin{align}
    \tau_n = \min_{\substack{U_n, \ket{\psi(0)} \\ \bra{\psi(0)}U_n\ket{\psi(0)}=0}} T_{\rm gate}(U_n)
\end{align}
is the duration of the gate that most rapidly evolves some initial state $\ket{\psi(0)}$ to an orthogonal state among all possible $n$-qubit operations $\{U_n\}$ on the given quantum computer. Here, $T_{\rm gate}(U_n)$ is the physical time required to implement the gate $U_n$.

By the universality of QSLs, these estimators have general applicability independent of the details of the Hamiltonian. In particular, it is valid for time-dependent Hamiltonians, which are common in practical quantum gate implementations. For example, in superconducting qubits, time-dependent drives are used to suppress leakage to higher levels for short gate times \cite{Rebentrost2009-xu,Motzoi2009-rw,Krantz2019-dw,Hyyppa2024-au}. Even in such time-dependent cases, the MT bound gives a rigorous lower bound for the time-averaged variance: $\Delta E_n \geq \Delta E_n^{\rm est}$. On the other hand, the ML bound assumes a time-independent Hamiltonian \cite{Okuyama2018-ii}, and thus $E_n^{\rm est}$ gives a lower bound on the effective Hamiltonian $H_{\rm eff}$ rather than the actual time-dependent Hamiltonian $H_n(t)$, as follows. 

For any unitary $U_n$ implemented in time $\tau_n$, one can always define an effective time-independent Hamiltonian $H_{\rm eff}$ satisfying $U_n = \exp(-i\tau_n/\hbar\, H_{\rm eff})$; by choosing an appropriate branch, the eigenvalues of $H_{\rm eff}$ can be taken as small nonnegative values as possible. For $H_{\rm eff}$, the ML bound applies exactly and thus our estimator bounds it: $E_{\rm eff}=\bra{\psi(0)}H_{\rm eff}\ket{\psi(0)} \geq E_n^{\rm est}$. The difference between $E_{\rm eff}$ and $E_n$ is analyzed by series expansion: specifically, the Magnus expansion for $H_{\rm eff}$ and the Dyson series for $U_n(t)$. Using these expansions, we can express both $E_{\rm eff}$ and $E_n$ in terms of $H_n(t)$ and commutators thereof regarding different times. As a result, we find that $E_{\rm eff}$ and $E_n$ coincide up to the first order with respect to $H_n(t)$ and the difference appears from the second order as
\begin{align}
    &|E_{\rm eff} - E_n|\nonumber\\& = \frac{1}{2\hbar\tau_n}\left|\int_{0<t_2<t_1<\tau_n}\bra{\psi(0)} [H_n(t_1), H_n(t_2)] \ket{\psi(0)} dt_1 dt_2\right|\nonumber\\&\quad + \mathcal{O}(H^3).\label{eq:energy_difference}
\end{align}
We provide the derivation in Appendix. From Eq.~\eqref{eq:energy_difference}, we expect that $E_{\rm eff}$ provides a good approximation of $E_n$ when these higher-order terms are negligible. The difference $|E_{\rm eff}-E_n|$ depends on the non-commutativity of $H_n(t)$ at different times: specifically, the non-commutativity between $H_n(t)$ and its time derivative $dH_n(t)/dt$. Therefore, if the driving field is designed to counteract the influence of the rapid changes of the Hamiltonian, as in the derivative removal by adiabatic gate approach \cite{Motzoi2009-rw}, the estimator is expected to be more accurate.

Next, we describe the way to extract the gate time $\tau_n$ from the job execution time of a black-box quantum computer. In typical cloud quantum computers, the job execution time $T_{\rm exec}$ returned to the user includes all components beyond quantum gate operations, such as device initialization, readout, and classical processing. Moreover, the time resolution is on the second scale, and one cannot directly read out nanosecond-scale single-gate durations. 

We resolve this issue with a simple but effective scaling method, gate-time amplification. Let $N_{\rm gate}$ be the number of occurrences of an $n$-qubit gate in a circuit and $N_{\rm shots}$ be the number of shots. Then the job execution time $T_{\rm exec}$ can be approximated as
\begin{align}
    T_{\rm exec} &\simeq N_{\rm circ} N_{\rm shots} \left(T_{\rm init} + N_{\rm gate} T_{\rm gate} + T_{\rm meas}\right)\nonumber\\
    &\quad+ \mathcal{T}(N_{\rm circ},N_{\rm shots})
\end{align}
where $T_{\rm init}$ and $T_{\rm meas}$ are the average times for state preparation and measurement, $T_{\rm gate}$ is the physical time for one target gate operation, $N_{\rm circ}$ is the number of circuits in the job, and $\mathcal{T}(\cdots)$ represents other overheads \cite{IBMUnknown-eb}. For the regime where $N_{\rm gate}$ is large, we expect that the contribution from gate operations dominates the change in execution time. Therefore, by varying $N_{\rm gate}$ while fixing the other parameters, one can estimate $T_{\rm gate}$ from the slope of $T_{\rm exec}$ with respect to $N_{\rm gate}$. For example, with $N_{\rm gate}=10^5$ and $N_{\rm shots}=10^3$, even a gate time of a few tens of nanoseconds contributes to the job time on the second scale, which is detectable with second-level resolution. Overall, our method, composed of the energy estimators and gate-time amplification, is generally applicable: even if a quantum computer does not explicitly return execution time, the user can measure the wall-clock time from job submission to completion.

\textit{Demonstrations---}We apply the above framework to IBM's superconducting quantum processor \texttt{ibm\_torino}, an instance of the ``Heron'' processor \cite{IBMUnknown-iw,McKay2023-tj}. The processor is based on transmon qubits and is accessible via the cloud. In the user interface, one specifies a set of quantum circuits executed as one job and the number of shots for each circuit; the measurement results and the job execution time $T_{\rm exec}$ are returned. The error rate of single-qubit gates is on the order of $10^{-4}$, and that of two-qubit gates is on the order of $10^{-3}$ \cite{IBMUnknown-du}. We note that IBM Quantum has been gradually restricting access to lower-level information, such as pulse-level control, qubit resonant frequencies, and anharmonicity \cite{IBMUnknown-to,IBMUnknown-du}. Therefore, the present black-box approach is becoming increasingly relevant.

\begin{figure}[t]
    \centering
    \includegraphics[width=0.7\linewidth]{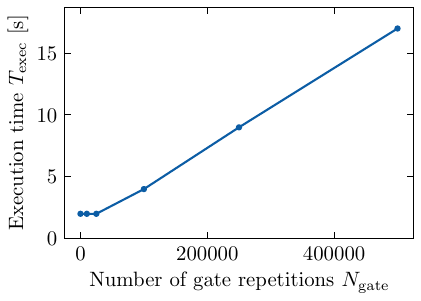}
    \caption{Example of gate-time amplification for the $X$ gate on \texttt{ibm\_torino}. The vertical axis shows the job execution time $T_{\rm exec}$ and the horizontal axis shows the number of repeated $X$ gates $N_{\rm gate}$ contained in each circuit. Each data point corresponds to a job consisting of a single circuit with $N_{\rm shots}=10^3$.}
    \label{fig:gate_time_amplification}
\end{figure}

First, we estimate the gate time of the X gate as an example of gate-time amplification. We construct a set of one-qubit circuits, each of which applies the $X$ gate consecutively $N_{\rm gate}$ times to the initial state $\ket{0}$. We fix the number of shots to $N_{\rm shots}=10^3$ for each circuit and vary $N_{\rm gate}$ from $0$ to $5\times 10^5$. For each $N_{\rm gate}$, we create a job consisting of a single circuit and record the job execution time $T_{\rm exec}$. Figure~\ref{fig:gate_time_amplification} shows the empirical values of $T_{\rm exec}$ as a function of $N_{\rm gate}$. We observe an approximately linear relation in the regime where $N_{\rm gate}$ is sufficiently large ($N_{\rm gate}\geq  10^5$). By performing linear regression using the data in this regime, we estimate the gate time per $X$ gate as 32 ns. This value agrees with a gate time based on vendor-released calibration information (32 ns) \cite{IBMUnknown-du}, supporting the validity of the gate-time amplification method. We note that this calibration information is only used here for validation; our method does not require such prior knowledge.

\begin{table}[t]
\centering
\caption{Estimated gate times for representative single-, two-, and three-qubit gates on \texttt{ibm\_torino}, and the corresponding estimated energy scale.}
\label{table:gate_time_and_energy}
\begin{tabular}{lccc}
\toprule
    & Gate & Gate time $T_{\rm gate}$ & \parbox{3cm}{Energy estimate \\ $E_n^{\rm est}, \Delta E_n^{\rm est}$}\\
\midrule
\multirow{3}{*}{\parbox{2cm}{Single-qubit\\($n=1$)}}
    & X & 32 ns & \multirow{3}{*}{$5.2 \times 10^{-27}$ J}\\
    & Y & 32 ns & \\
    & Z & 0 ns &\\
\midrule
\multirow{3}{*}{\parbox{2cm}{Two-qubit\\($n=2$)}}
    & CZ   & 70 ns  & \multirow{3}{*}{$2.4\times 10^{-27}$ J} \\
    & CNOT & 140 ns & \\
    & iSWAP&  220 ns &\\
\midrule
\multirow{3}{*}{\parbox{2cm}{Three-qubit\\($n=3$)}}
    & Toffoli & 1500 ns & \multirow{3}{*}{$3\times 10^{-28}$ J} \\
    & iToffoli & 500 ns & \\
    & CCZ     & 1600 ns & \\
\bottomrule
\end{tabular}
\end{table}

By applying the same procedure to various gates, we estimate the shortest gate time $\tau_n$ required to orthogonalize a state. For single-qubit gates, we consider $\{X,Y,Z\}$, for two-qubit gates $\{\mathrm{CZ}, \mathrm{CNOT}, \mathrm{iSWAP}\}$, and for three-qubit gates $\{\mathrm{Toffoli}, \mathrm{iToffoli}, \mathrm{CCZ}\}$ \cite{Nielsen2010-tb}. These gates evolve at least one initial state to an orthogonal state. Even if the fastest orthogonalizing gate were implemented as a gate outside these sets, the estimated times of these gates still serve as valid lower bounds. Since the time of each gate does not depend on the input state, we fix the initial state to $\ket{0}^{\otimes n}$. The results are summarized in Table~\ref{table:gate_time_and_energy}. For the $Z$ gate, the estimate suggests an effectively instantaneous execution, consistent with the fact that in superconducting qubits, the $Z$ gate is typically implemented as a virtual gate via a phase-frame update without a physical pulse. We note that the gate times generally depend on the connectivity of qubits. For example, the Toffoli gate on linearly connected three qubits takes longer than that on fully connected three qubits: in addition, for the linear connectivity, the time varies depending on which qubit is used as the target. In Table~\ref{table:gate_time_and_energy}, we show the shortest gate times among different qubit choices.

Among these gate times, we take the smallest as $\tau_n$ and substitute it into Eq.~\eqref{eq:estimator_multi} to estimate the energy scale of each $n$-qubit Hamiltonian. The results are summarized in Table~\ref{table:gate_time_and_energy}. For single-qubit gates, substituting $\tau_1=32$ ns into Eq.~\eqref{eq:estimator_multi} yields
\begin{align}
    E_1^{\rm est}=\Delta E_1^{\rm est}= 5.2\times 10^{-27}\,{\rm J}
\end{align}
Similarly, we obtain $E_2^{\rm est}=\Delta E_2^{\rm est}= 2.4\times 10^{-27}$ J for two-qubit gates and $E_3^{\rm est}=\Delta E_3^{\rm est}= 3\times 10^{-28}$ J for three-qubit gates.

\textit{Discussion---}We first discuss the single-qubit-gate result. As a reference, we consider typical driving strengths in superconducting transmon qubits. In the literature, Rabi frequencies $\Omega/(2\pi)$ for single-qubit gates are reported to be around $10$--$100$ MHz (at their peak) \cite{Krinner2019-ke,Krantz2019-dw,Gao2021-jc,Shirai2023-jw,Xu2025-kx}. For simplicity, we assume that the X gate is implemented with a square pulse and the initial state is $\ket{0}$. Setting the ground state energy to zero, the energy expectation and variance during the gate is $ 3.3\times 10^{-27}$--$3.3\times 10^{-26}$ J, which is consistent with our estimate $5.4\times 10^{-27}$ J. This agreement supports the effectiveness of our method. Specifically, although our estimators provide only lower bounds on the actual energy, these bounds are considered close to the true values. In other words, the success of our method suggests that  the present quantum processor operates near the QSL for single-qubit gates. For such a system where QSL serves as a design principle \cite{Caneva2009-zf}, only the time information is sufficient to infer the energy lower bound by our approach. We note that the influence of the error on the gate operation is negligible, as discussed in Appendix.

The results for two-qubit gates further demonstrate the effectiveness of our method. Implementations of two-qubit gates are more diverse than those of single-qubit gates \cite{Gao2021-jc}. There are multiple design choices, including the type of two-qubit gate (e.g., CZ, iSWAP, and cross-resonance), the coupler architecture, and the pulse shape \cite{Sheldon2016-me,Chow2011-yr,Majer2007-cl,Hong2020-dy,Caldwell2018-xf,Strauch2003-yv,DiCarlo2009-el,Rol2019-oo,Sung2021-ky,Krantz2019-dw}. Even when restricting attention to CZ gates, which are the native two-qubit gates on \texttt{ibm\_torino}, various implementations have been proposed \cite{Barends2013-gg,Reagor2018-wo,Caldwell2018-xf,Strauch2003-yv,Sung2021-ky}. Our method is applicable without any prior knowledge about such an internal implementation. To evaluate the validity of our estimate, we refer to the effective ZZ coupling strength in the literature, ranging between 1--20 MHz \cite{Li2020-pc,Shirai2023-jw,Howard2023-hb,Marxer2023-nu}. Assuming time-independent $\sigma_z\otimes\sigma_z$ coupling, the corresponding energy expectation and variance is around $3.3\times10^{-28}$--$6.6\times 10^{-27}$ J, which is also consistent with our estimate.

Finally, for three-qubit gates, we obtain a lower bound of order $10^{-28}$ J, which is about one order of magnitude smaller than those for single- and two-qubit gates.  On \texttt{ibm\_torino}, three-qubit gates are not natively implemented but are decomposed into single- and two-qubit gates, as is common in many quantum computing platforms \cite{Preskill2018-ug,Kusyk2021-rk}. Therefore, the estimated value should be interpreted as a lower bound on the time-averaged energy of the overall three-qubit dynamics, which sets a reference for the black-box three-qubit operations. Moreover, three-qubit gates are reported in the superconducting qubit experiments \cite{Liu2025-dh}, supporting the potential relevance of our approach.

\textit{Conclusion---}We proposed a method to infer the internal energy scales of black-box quantum computers using the QSL and demonstrated its effectiveness on IBM's superconducting quantum processor. The present method provides an example of using time--energy uncertainty as an inference tool and is expected to be extendable to estimation of broader quantum-thermodynamic parameters, including energy gaps, temperature, or other conjugate observables by utilizing appropriate QSLs or uncertainty relations. Moreover, the method is applicable not only to quantum computers but also to other quantum systems in which the Hamiltonian is inaccessible while operation durations are measurable. It provides a noninvasive inference perspective based on uncertainty relations and can be a practical protocol in quantum thermodynamics and quantum information processing.

\vspace{1em}
\begin{acknowledgments}
  \noindent
  \textit{Acknowledgments---}We acknowledge the use of IBM Quantum services for this work. The views expressed are those of the authors, and do not reflect the official policy or position of IBM or the IBM Quantum team.
\end{acknowledgments}

\par
\appendix
\section{Appendix}
\subsection{Analysis of the effective Hamiltonian}
We analyze the estimator $E_n^{\rm est}$ for the time-dependent Hamiltonian case in detail. For notational simplicity, we omit the subscript $n$ in this section. We consider a unitary operator $U$ implemented in time $\tau$ by a time-dependent Hamiltonian $H(t)$. As in the main text, the effective time-independent Hamiltonian $H_{\rm eff}$ is defined as $U = \exp(-i\tau/\hbar\, H_{\rm eff})$. This $H_{\rm eff}$ can be expressed in terms of $H(t)$ using the Magnus expansion:
\begin{align}
    &H_{\rm eff} = \frac{1}{\tau}\int_0^\tau H(t_1) dt_1\nonumber\\ &+ \frac{1}{2i\hbar\tau} \int_{0<t_2<t_1<\tau} [H(t_1), H(t_2)] dt_1 dt_2 + \mathcal{O}(H^3).
\end{align}
The energy expectation with respect to the initial state $\ket{\psi(0)}$ is then given by
\begin{align}
    E_{\rm eff} &= \bra{\psi(0)} H_{\rm eff} \ket{\psi(0)} \nonumber\\
    &= \frac{1}{\tau}\int_0^\tau \bra{\psi(0)} H(t) \ket{\psi(0)} dt\nonumber\\ &\quad+ \frac{1}{2i\hbar\tau} \int_{0<t_2<t_1<\tau} \bra{\psi(0)} [H(t_1), H(t_2)] \ket{\psi(0)} dt_1 dt_2 \nonumber\\&\quad+ \mathcal{O}(H^3).\label{eq:appendix:effective_energy}
\end{align}

On the other hand, for the time-averaged energy expectation $E$, we use the Dyson series to express $U(t)$ as
\begin{align}
    &U(t) = \mathbb{I} - \frac{i}{\hbar} \int_0^t H(t_1) dt_1\nonumber\\& + \left(-\frac{i}{\hbar}\right)^2 \int_{0<t_2<t_1<t} H(t_1) H(t_2) dt_1 dt_2 + \mathcal{O}(H^3).
\end{align}
Then, we can expand $U^\dagger(t) H(t) U(t)$ as
\begin{align}
    U^\dagger(t) H(t) U(t) &= H(t) + \frac{i}{\hbar} \int_0^t [H(t_1), H(t)] dt_1 + \mathcal{O}(H^3).
\end{align}
Therefore, the time-averaged energy expectation is obtained by integrating the both sides over time:
\begin{align}
    &E = \frac{1}{\tau} \int_0^\tau \bra{\psi(0)} U^\dagger(t) H(t) U(t) \ket{\psi(0)} dt \nonumber\\
    &= \frac{1}{\tau} \int_0^\tau \bra{\psi(0)} H(t) \ket{\psi(0)} dt\nonumber\\& + \frac{1}{i\hbar\tau} \int_{0<t_2<t_1<\tau} \bra{\psi(0)} [H(t_1), H(t_2)] \ket{\psi(0)} dt_1 dt_2 + \mathcal{O}(H^3).\label{eq:appendix:time_averaged_energy}
\end{align}
By comparing Eqs.~\eqref{eq:appendix:effective_energy} and \eqref{eq:appendix:time_averaged_energy}, we find that they coincide up to the first order and the difference appears from the second order as
\begin{align}
    &|E_{\rm eff} - E|\nonumber\\& = \frac{1}{2\hbar\tau}\left|\int_{0<t_2<t_1<\tau}\bra{\psi(0)} [H(t_1), H(t_2)] \ket{\psi(0)} dt_1 dt_2\right|\nonumber\\&\quad + \mathcal{O}(H^3).
\end{align}

\subsection{Effect of gate errors on the estimation}
In practical quantum computers, gate operations are subject to errors. Here, we analyze the effect of such errors on the QSL, which is the basis of our estimation method. For simplicity, we consider the coherent error. Let the initial state be $\ket{\psi(0)}$ and the ideal final state after the gate operation be $\ket{\psi_\perp}$, which is orthogonal to $\ket{\psi(0)}$. Due to the coherent error, the actual final state becomes $\ket{\psi_{\rm err}}$, which deviates from $\ket{\psi_\perp}$. The error is quantified by the fidelity between the ideal and actual final states: $|\braket{\psi_\perp}{\psi_{\rm err}}|^2 = 1 -  \epsilon$ with a small error parameter $\epsilon > 0$. The Bures angle $\mathcal{L}$ between the initial and actual final states is given by
\begin{align}
    \mathcal{L} = \arccos|\braket{\psi(0)}{\psi_{\rm err}}| = \arccos\sqrt{\epsilon}\sim \frac{\pi}{2} - \sqrt{\epsilon}.
\end{align}
Then, the MT bound for this case is expressed as
\begin{align}
    T_{\rm QSL}^{\rm MT} = \frac{\hbar \mathcal{L}}{\Delta E}\sim \frac{\pi\hbar}{2\Delta E}\qty(1 - \frac{2\sqrt{\epsilon}}{\pi})
\end{align}
Therefore small coherent errors do not change the leading scaling of $T_{\rm QSL}^{\rm MT}$, while they induce an additive correction proportional to $\sqrt{\epsilon}$. In conclusion, for high-fidelity gates with small $\epsilon$, the effect of gate errors on our energy estimation method is negligible if the leading order is concerned.

\end{document}